\documentclass[12pt]{article}
\usepackage{amsmath,amssymb}
 \newcommand{\eqr}[1]{(\ref{#1})}
 \newcommand{\ds}{\displaystyle}
\newtheorem{theorem}{Theorem}
\numberwithin{equation}{section}

\newcounter{saveeqn}%
\newcommand{\alpheqn}{\setcounter{saveeqn}{\value{equation}}%
\stepcounter{saveeqn}\setcounter{equation}{0}%
\renewcommand{\theequation}
{\mbox{\arabic{section}.\arabic{saveeqn}\alph{equation}}}}%
\newcommand{\reseteqn}{\setcounter{equation}{\value{saveeqn}}%
\renewcommand{\theequation}{\arabic{section}.\arabic{equation}}}%

\begin{document}
\today

\begin{center}
\bigskip

{\bf Point Symmetries of Generalized Toda Field Theories II\\
Applications of the Symmetries}

\bigskip L. Martina  \footnote{E-mail:
martina@le.infn.it} 

{\it Dipartimento di Fisica dell'Universit\`a and INFN
Sezione di 
 Lecce, C.P. 193, 73100 Lecce, Italy}
 
S. Lafortune\footnote{E-mail: lafortus@CRM.UMontreal.CA}  and P. 
Winternitz\footnote{E-mail:
wintern@CRM.UMontreal.CA} 

{\it Centre des Recherches Math\'{e}matiques, Universit\'{e} de Montr\'{e}al,
C.P. 6128, Succ. Centre-Ville, Montr\'{e}al, Qu\'ebec, H3C 3J7, Canada }

\end{center}

\vspace{2in}

\medskip Abstract \qquad {\it The Lie symmetries of a large class of generalized 
Toda 
field theories are studied and used to perform symmetry reduction. Reductions 
lead to 
generalized Toda lattices on one hand, to periodic systems on the other. 
Boundary 
conditions are introduced to reduce theories on an infinite lattice to those on 
semi-infinite, or finite ones.}

\newpage 

\section{Introduction\protect\bigskip}

In a recent article \cite{1} we determined the Lie point symmetries of
a class of equations that we called ``generalized Toda field
theories''. They all involved various types of exponentials and had
the form
\begin{equation}
u_{n,xy}= F_{n},\hspace{1cm} F_{n}=  \sum_{m=  n-n_{1}}^{n+n_{2}} K_{nm} \exp 
\left( {{\sum_{l=  m-n_{3}}^{m+n_{4}} H_{ml} u_{l}}}\right),  \label{1.1}
\end{equation}
where $K$ and $H$ are some real constant matrices and $n_{1},\dots
,n_{4}$ 
are   
some
finite non-negative integers. 
The symmetries were obtained independently for three different ranges
of the variable $n$. Namely, we considered the infinite case
$-\infty<n<\infty$, the semi-infinite one, $1\leq n<\infty$, and also the
finite case, $1\leq n\leq N<\infty$.

In the infinite and semi-infinite cases eq. \eqr{1.1} was treated as a
differential-difference equation. If the range of $n$ is finite,
eq. \eqr{1.1} is simply a system of $N$ differential equations for $N$
unknowns $u_n$.

A sizable literature exists on symmetries of difference equations
\cite{1}-\cite{18}. Different approaches differ in their treatment of
independent variables and also in the degree of generalization of the
concept of ``point symmetries'' that is involved in passing from
differential equations to difference ones.

In Ref. \cite{1} we adopted the ``differential equation'' approach
proposed earlier \cite{4}. Essentially, eq. \eqr{1.1} was treated as a
system of infinitely many differential equations for infinitely many
fields $u_n(x,y)$ with $-\infty<n<\infty$, or $1\leq n<\infty$,
respectively.

The purpose of this article is to investigate applications of the
obtained Lie point symmetries. In particular we shall show how one can
perform various types of symmetry reduction, using these
symmetries. The reductions will be from infinite systems to
semi-infinite, finite or periodic ones. We will also consider
reductions of the number of independent variables, both continuous and
discrete. Each time a reduction is performed the resulting equations
will inherit some of the symmetries of the original system. We will
show how to obtain the ``inherited'' symmetry group and will compare it
with the entire Lie point symmetry group of the reduced system.

The symmetries are point ones, in the sense that we assume that the
symmetry algebra, i.e. the Lie algebra of the symmetry group, is
realized by vector fields of the form
\begin{equation}
\hat{v}=\xi(x,y,\{u_k\})\partial_x+\eta(x,y,\{u_k\})\partial_y+
\sum_{j}\phi_j(x,y,\{u_k\})\partial_{u_j},
\label{1.2}
\end{equation}
where $\{u_k\}$ denotes the set of all fields (with $k$ in an
infinite, semi-infinite, or finite range, repectively). The summation
in eq. \eqr{1.2} is also over the appropriate range of values of $j$.

Thus, if eq. \eqr{1.1} is considered as an equation on a lattice, the
coefficients of the vector field $\hat{v}$ can depend on the values of
$u_k$ at all points of the lattice.

Whatever the range of the discrete {\it variable} $n$, the range of
the {\it interaction} $F_n$ in eq. \eqr{1.1} is assumed to be finite,
i.e. the integers $n_1,\dots, n_4$ are finite. Obviously, this allows
much more general interactions than nearest neighbour ones. Indeed
eq. \eqr{1.1} is general enough to include all Toda field theories
that, to our knowledge, occur in the literature, be they integrable or
not.

Thus, if we put $H_{n,n-1}=-H_{nn}=1$ and $K_{nn}=-K_{nn+1}=1$, and all
the other components of $H$ and $K$ equal
to zero, we obtain the ``two-dimensional Toda lattice''
\begin{equation}
u_{n,xy}=e^{u_{n-1}-u_n}-e^{u_n-u_{n+1}},\quad
-\infty<n<\infty
\label{1.3}
\end{equation}
originally introduced by Mikhailov \cite{19} and Fordy and Gibbons
\cite{20} and studied further in Ref. \cite{21}.

Other well known Toda systems (with nearest neighbour interactions)
are obtained if we set $H_{ml}=\delta_{ml}$
\begin{equation}
u_{n,xy}=   \sum_{m=  n-n_{1}}^{n+n_{2}} K_{nm} \exp 
u_{m},  \label{1.4}
\end{equation}
or vice versa $K_{nm}=\delta_{nm}$
\begin{equation}
u_{n,xy}=    \exp 
{\sum_{l=n-n_3}^{n+n_4}H_{nl} u_{l}}.  \label{1.5}
\end{equation}
All of these systems have been studied from the point of view of
their integrability and solutions \cite{19}-\cite{29}, usually for a
finite number of fields and usually for $K$, or respectively $H$, the
Cartan matrix of a simple, or affine, Lie algebra.

We shall call the theories \eqr{1.3}, \eqr{1.4} and \eqr{1.5} type I, II and III, 
respectively. We shall use the $n\rightarrow \infty$ generalization of an
${\mbox sl}(n+1,\mathbb{C})$ Cartan matrix, i.e. put
$K_{n-1n}=K_{n+1n}=-1$, $K_{nn}=2$, $K_{nm}=0$ for $m\neq n,\;n\pm 1$
in eq. \eqr{1.4} for $-\infty<n<\infty$. The same will be chosen for
$H$ in eq. \eqr{1.5}.

We note that the equations \eqr{1.4} and \eqr{1.5} are equivalent if
the range of $n$ is finite and the matrices $H$ and $K$ are
invertible. Indeed, if $u_n(x,y)$ satisfies eq. \eqr{1.4}, then $w=K^{-1}u$ 
satisfies eq. \eqr{1.5} 
with 
$H=K$, and vice-versa. For $-\infty<n<\infty$, or $1\leq n<\infty$, this is no
longer the case and their symmetry groups, in general, are different.

The original Toda lattice \cite{30,31} is obtained from eq. \eqr{1.3}
by symmetry reduction, using translational invariance, i.e. looking
for solutions invariant under translations generated by
\begin{equation}
P=\partial_x-\partial_y,\quad u_n(x,y)=u_n(t),\quad t=x+y.
\label{1.6}
\end{equation}
Since the matrices $H$ and $K$ are constant, the same reduction takes
eq. \eqr{1.1} into the generalized Toda lattice
\begin{equation}
\ddot{u}_{n}= F_{n},\hspace{1cm} F_{n}=  \sum_{m=  n-n_{1}}^{n+n_{2}} K_{nm} 
\exp 
\left( {{\sum_{l=  m-n_{3}}^{m+n_{4}} H_{ml} u_{l}}}\right).  \label{1.7}
\end{equation}

In Section 2, we study the case where the range of $n$ in eq. \eqr{1.1} is 
infinite. We give the 
symmetry 
algebra $L_f$ of the general Toda field theory \eqr{1.1} as calculated in 
\cite{1}. The same 
results 
concerning the particular theories of types I, II and III are also 
given. Eq. 
\eqr{1.7} inherits a subgroup of the symmetry group of eq. \eqr{1.1}. The Lie 
algebra 
$L_0$ of the ``inherited symmetry group'' will be the normalizer of the vector 
field $P=\partial_x-\partial_y$ in $L_f$
\begin{equation}
{\mbox{nor}}_{L_f}P=\{x\in L_f | [x,P]=\lambda P\}, \;\lambda\in\mathbb{R}.
\label{nor}
\end{equation}
Commuting $P$ with a general element of the algebra 
$L_f$, imposing the above normalizer condition and letting the resulting vector 
field act on functions of $t$ and $u_n$ we obtain the inherited symmetry algebra 
of the Toda 
lattice. We establish that the entire symmetry algebra of the generalized Toda 
lattice \eqr{1.7} coincides with the one inherited from $L_f$. From this general 
result, the 
symmetry algebras of the Toda 
lattices corresponding to the types I, II and III are 
obtained explicitely.

In Section 3, we restrict the range of the discrete variable to be $1\leq
n< \infty$. This means that starting from some value $n_0$ of $n$ the
eq. \eqr{1.1} and \eqr{1.7} will be the same as in the infinite case
but for $1\leq n < n_0$ they will be modified. Their actual form
will depend on the choice of the matrices $K$ and $H$.
Our procedure will be the same in all cases. We start from the already
established symmetry algebra in the case of infinitely many fields. Its
prolongation will annihilate all the equations in the system that are
not modified by the boundary conditions. We apply the prolonged vector
field to the equations for $1\leq n<n_0$ and require that these
equations also be annihilated on the solution set.  This will provide us with a 
subgroup of the 
original
symmetry group which is the symmetry group of the semi-infinite Toda
field theory. It must then be checked whether this ``inherited''
symmetry group is indeed the entire symmetry group of the corresponding
semi-infinite system, or only a subgroup of it.

In Section 4, we further restrict the range of the variable $n$ and consider 
Toda 
field theories 
and Toda lattices 
with a finite number of fields, $1\leq n \leq N$. We shall proceed as in the 
semi-infinite 
case, that is 
start from the symmetries  of Toda field 
theories, or 
Toda lattices, 
with $-\infty < n<\infty$. We then impose that the general symmetry generator 
should also 
annihilate the 
modified equations at the beginning and end of the chain. This is equivalent to 
starting 
from a 
semi-infinite Toda theory and requiring that the symmetry generator should also 
annihilate 
the equation for 
$n=N$.

In Section 5, we view the relation between infinite and periodic (generalized) 
Toda systems as 
symmetry 
reduction. Indeed, let us consider the Toda field theory \eqr{1.1} 
for $n \in 
\mathbb{Z}$ and its symmetry algebra. In order to be able to impose 
periodicity
\begin{equation}
u_{n+N}(x,y)=u_n(x,y),
\label{per}
\end{equation}
 the symmetry 
algebra 
 of the generalized Toda field theories or of the generalized 
Toda 
lattices, should be enlarged by adding the operator 
\begin{equation}
\hat{N}=\partial_{n}
\label{N}
\end{equation}
to the corresponding symmetry algebra, whenever possible. This operator generates 
shifts 
(``translations'' in 
$n$), though the corresponding translation group parameter must be by an 
integer, 
$\tilde{n}=n+\lambda,\quad\lambda\in \mathbb{Z}$. Imposing periodicity, as in 
eq. 
\eqr{per}, means that $\hat{N}$ must be removed from the symmetry algebra. The 
symmetry 
algebra of the periodic system, inherited from the infinite one, will be the 
normalizer of 
$\hat{N}$ in the original algebra. The operator $\hat{N}$ of eq. \eqr{N} can 
also be used to reduce 
differential equations on lattices to differential-delay equations \cite{4}.

In section 6, we study a further application of the symmetry group. We look at symmetry 
reductions 
involving both discrete and continuous variables.

\section{Symmetries of infinite generalized Toda field theories and
  lattices}

\subsection{General results}

Let us now consider the differential-difference eq. \eqr{1.1} and for 
$-\infty <n<\infty$ impose that the matrices $H$ and $K$ are band matrices with 
finite bands of constant width:
\begin{equation}
H_{n m}=  H_{n,\, n+\sigma }=  \left\{  
\begin{array}{l}
{h_{\sigma }(n)\;\;\sigma \in [p_{1},p_{2}]} \\  
{0\qquad\; \sigma \not\!{\in}\, [p_{1},p_{2}]}
\end{array}
\right. {,\; {h_{p_{1}}(n)\neq 0,\;\;h_{p_{2}}(n)\neq 0, \quad p_1\leq p_2.}} 
 \label{2.3}
\end{equation}
Similarly,  
\begin{equation}
K_{n m}=  K_{m+\sigma, m}=  \left\{  
\begin{array}{l}
{k_{\sigma }(m)\;\;\sigma \in [q_{1},q_{2}]} \\  
{0\;\;\;\;\;\;\;\;\;\sigma \not\!{\in}\, [q_{1},q_{2}]}
\end{array}
\right. {,\; k_{q_{1}}(m)\neq 0,\;\;k_{q_{2}}(m)\neq 0, \quad q_1\leq q_2. }  
\label{2.4}
\end{equation}

We introduce the quantity $\rho_n$ which, as a function of $n$, is any 
particular solution of the 
inhomogeneous linear 
finite-difference equation
\begin{equation}
\sum_{\sigma =  p_{1}}^{p_{2}}h_{\sigma }(n)\rho_{\sigma +n} =  1
\label{2.5}
\end{equation}
and the quantities $\psi^j_n$, $j=1, \dots, p_2-p_1$ and $\phi^{l}_{m}$, $l=1, 
\dots, 
q_2-q_1$, which are, respectively, linearly independent 
solutions of the homogeneous 
linear equations
\begin{equation}
\sum_{\sigma =  p_{1}}^{p_{2}}h_{\sigma }(n)\ \psi _{\sigma +n} =  0,\quad
\sum_{\sigma =  q_{1}}^{q_{2}}k_{\sigma }\left( m \right) \ \phi _{\sigma   
+m}=  0.
\label{2.6}
\end{equation}

Without proof we present the following result \cite{1}.

\begin{theorem}
The symmetry algebra of the infinite Toda field theory \eqr{1.1} with 
$-\infty<n<\infty$ and $H$ and $K$ satisfying \eqr{2.3} and \eqr{2.4} has a 
basis consisting of the following vector fields 
\alpheqn
\begin{equation}
X=  \xi (x)\partial_{x}+\eta(y)\partial_y - 
(\xi_{x}+\eta_y) \sum_{n=-\infty}^\infty \rho_{n}\partial_{u_{n}},  
\label{2.2a}
\end{equation}
\begin{equation}
V_{j}=  (r_j(x)+s_j(y)) \sum_{n=-\infty}^\infty\psi   
_{n}^{j}\partial
_{u_{n}},\label{2.2b}
\end{equation}
\begin{equation}
Z_{jl}=  \left( \sum_{m=-\infty}^\infty\phi _{m}^{l}u_{m}\right) \left(   
\sum_{n=-\infty}^\infty\psi
_{n}^{j}\partial _{u_{n}}\right),  \label{2.2c}
\end{equation}
$$
 j=  1,\dots
,p_{2}-p_{1},\quad l=  1,\ldots ,q_{2}-q_{1}.
$$
\reseteqn
\addtocounter{equation}{-1}\refstepcounter{equation}\label{2.2}
The functions $\xi (x),\,\eta (y),\, r_j\left( x\right) $ and $s_j\left(   
y\right) $
are arbitrary and $\rho_n$, $\psi_n^j$ and $\phi_n^l$ are defined in
eq. \eqr{2.5} and \eqr{2.6}, respectively.
\end{theorem}

The operator $X$ reflects the fact that the generalized Toda 
field 
theories with $-\infty <n<\infty$ are always conformally invariant, be they 
integrable or not. If 
we have 
$p_2-p_1\geq 1$ the theory is invariant under gauge transformations. 
If we have 
$q_2-q_1\geq 1$ a further type of gauge invariance exists, represented by the 
operator $Z_{jl}$. If $Z_{jl}$ is absent the gauge group is abelian, 
otherwise it is non abelian. The commutation relations are given
elsewhere \cite{1}.

The generalized Toda Lattice \eqr{1.7} is obtained from the generalized Toda 
field theory 
by symmetry reduction. Indeed, let us reduce by the translation generator 
$P=\partial_x-\partial_y$ as in eq. \eqr{1.6}. Eq. \eqr{1.1} 
reduces to eq. \eqr{1.7}. We calculate the symmetry subgroup of the symmetry 
group of eq. \eqr{1.1} 
inherited 
by eq. \eqr{1.7} using the procedure explained in the Introduction. We obtain 
the following result.

\begin{theorem} The basis for the 
inherited symmetry algebra for the infinite Toda lattice \eqr{1.7}
with $H$ 
and 
$K$ satisfying \eqr{2.3} and \eqr{2.4} is given by
\begin{equation}
\begin{array}{cc}
\ds{P=\partial_t,}&\ds{D=t\partial_t-2\sum_{n=-\infty}^{\infty}\rho_n\partial_{u
_n}
}\\ \\
\ds{U_j=\sum_{n=-\infty}^{\infty}{\psi_{n}}^j\partial_{u_n},}&
\ds{W_j=t\sum_{n=-\infty}^{\infty}{\psi_{n}}^j\partial_{u_n}}
\end{array}
\label{2.8}
\end{equation}
and $Z_{jl}$ as in eq. \eqr{2.2c}. The inherited algebra coincides with 
the actual symmetry algebra of eq. \eqr{1.7}.
\end{theorem}

Just as in the case of differential equations, there is no guarantee that the 
symmetry algebra inherited from the original equation is the entire symmetry 
algebra of the reduced equation. However, in this case, a direct calculation 
of the symmetry algebra of the generalized Toda lattice shows that its symmetry 
algebra is indeed given by eq. \eqr{2.8}.

\subsection{Special cases}

{\noindent}{\bf A. The type I system}

We have $p_2-p_1=q_2-q_1=1$, and obtain
\begin{equation}
\psi_m=\phi_m=1,\quad \rho_n=-n
\label{2.9}
\end{equation}
in eqs. \eqr{2.2} and \eqr{2.8}. Thus, \eqr{2.2a} is present, as are
\eqr{2.2b} and \eqr{2.2c}. The labels $j=l=1$ can be dropped.

{\noindent}{\bf B. The type II system with $K_{n-1n}=K_{n+1n}=-1$, 
$K_{nn}=2$}

We have $p_2=p_1=0$, $q_2-q_1=1$ and hence
\begin{equation}
\psi_m=0,\quad \phi_m=an+b,\quad\rho_n=1,
\label{2.10}
\end{equation}
with $a$ and $b$ constant.
Conformal invariance \eqr{2.2a} is present but there are no gauge
transformations \eqr{2.2b}, nor \eqr{2.2c}.

{\noindent}{\bf C. The type III system  with
  $H_{n-1n}=H_{n+1n}=-1$,
$H_{nn}=2$}

We have $p_2-p_1=2$, $q_2-q_1=0$ and 
\begin{equation}
\psi_n^1=1,\quad\psi_n^2=n,\quad\phi_m=0,\quad\rho_n=-\frac{1}{2}n^2.
\end{equation}
We have conformal invariance \eqr{2.2a}, two operators $V_1$ and
$V_2$, no $Z$. 

The symmetries of the corresponding generalized Toda lattices are
obtained by putting the above values of $\rho_n$, $\psi_m$ and
$\phi_m$ into eq. \eqr{2.8}.

\section{Reduction to semi-infinite theories}

We will now calculate the symmetry groups of semi-infinite theories inherited 
from infinite Toda 
systems using 
the procedure explained in the Introduction.

Rather than impose general and arbitrary boundary conditions on the
matrices $K$ and $H$ of eq. \eqr{1.1} we shall consider several special cases 
suggested by Lie
group theory, that already occured in the literature (usually for $n$
varying in a finite range).

\subsection{The semi-infinite field theories of type I
  related to simple Lie algebras}

Let us consider a semi-infinite generalization of the system \eqr{1.3}.
 The field equation is given by
\begin{equation}
{\bf U}_{xy}  = \mu^2  
\sum_{i 
=1}^{N}\frac{\mbox{\boldmath$\alpha$}_{i}}{\mbox{\boldmath$\alpha$}_{i}^2}
\exp(\mbox{\boldmath$\alpha$}_{i} \cdot {\bf U}), \label{MFG}
\end{equation}
where ${\bf U} =  ( u_{1}, \dots, u_N) $
is an $N$-tuple of real fields and $( \mbox{\boldmath$\alpha$}_{1},\dots,
\mbox{\boldmath$\alpha$}_{N} )$ denote
the simple roots of some finite simple Lie algebra of rank $N$. 
If this Lie algebra is ${\mbox
  sl}(N+1,\mathbb{R})$ one obtains the usual finite Toda field theory. Instead 
of
this we shall consider the Toda field theory \eqr{1.3} and let the Lie
algebra root system run through semi-infinite extensions of the
classical Cartan Lie algebras $A_N$, $B_N$, $C_N$ and $D_N$ with
$N\rightarrow \infty$ (in one direction).

For the system \eqr{1.3} the symmetry algebra \eqr{2.2} reduces to
\alpheqn
\begin{eqnarray}
&\ds{X=\xi(x)\partial _{x}+\eta(y)\partial_{y}+(\xi
  _{x}+\eta_y) \sum_{n=-\infty}^\infty n\partial_{u_{n}},}&\label{3.2a}
 \\
&\ds{V=    (\beta (x)+\gamma(y)) \sum_{n=-\infty}^\infty\partial_{u_{n}},}& 
\label{3.2b}\\
&\ds{Z=   \bigg{(} \sum_{m=-\infty}^\infty u_{m}\bigg{)} \bigg{(}
   \sum_{n=-\infty}^\infty\partial_{u_{n}}\bigg{)}.}&
\label{3.2c}
\end{eqnarray}
\reseteqn
\addtocounter{equation}{-1}\refstepcounter{equation}\label{3.2}

For the ordinary Toda lattice theory obtained by replacing $u_{n,xy}$
by $u_{n,tt}$ in eq. \eqr{1.3}, the symmetry algebra \eqr{2.8} reduces
to
\alpheqn
\begin{eqnarray}
&\ds{P=\partial _{t},\quad
D=t\partial_{t}+2\sum_{n=-\infty}^{\infty}n\partial_{u_{n}},}&
 \label{3.3a} \\
&\ds{U= \sum_{n=-\infty}^\infty\partial
_{u_{n}},\quad W=  
 t  \sum_{n=-\infty}^\infty\partial _{u_{n}},}& \label{3.3b}
\end{eqnarray}
\reseteqn
\addtocounter{equation}{-1}\refstepcounter{equation}\label{3.3}
and $Z$ as in \eqr{3.2c}.

We now turn to the semi-infinite case. For each algebra we show the
modified equations (for $n<n_0$). Those not shown coincide with
eq. \eqr{1.3}.

\noindent{\bf 1. The Cartan $A$ series}  

We have 
\begin{equation}
u_{1,xy}=-e^{u_1-u_2}.
\label{3.4}
\end{equation}

Taking a general element of the algebra \eqr{3.2}, applying its
prolongation to eq. \eqr{3.4} and taking into account that
\begin{equation}
\bigg{(} \sum_{i=1}^\infty u_i\bigg{)}_{xy}=0
\label{3.5}
\end{equation}
we find that the entire symmetry algebra \eqr{3.2} leaves
eq. \eqr{3.4} invariant. Hence the symmetry algebras in the semi-infinite and
infinite cases coincide, though in the semi-infinite case all summations
are for $1\leq n<\infty$. Eq. \eqr{3.5} was used to show that the
$Z$ symmetry also survives.

The same is true for the Toda lattice equations in this case. Namely,
the symmetry groups are the same in the infinite and semi-infinite
cases.

\noindent{\bf 2. The Cartan $B$ series}   

In this case, the first equation is
\begin{equation}
u_{1,xy}=e^{-u_1}-e^{u_1-u_2}.
\label{3.6}
\end{equation}
Applying the
same procedure, we find that conformal invariance remains and is
realized as in eq. \eqr{3.2a}. The gauge symmetries \eqr{3.2b} and
\eqr{3.2c} do not survive.

For the Toda lattice, the only surviving symmetries are the
translation $P$ and dilation $D$ as in eq. \eqr{3.3a}.

\noindent{\bf 3. The Cartan $C$ series}

We have
\begin{equation}
u_{1,xy}=-e^{u_1-u_2}+2e^{-2u_1}.
\label{3.7}
\end{equation}
In this case the
gauge symmetries combine together with the conformal ones. The
presence of eq. \eqr{3.7} excludes invariance under the $Z$
transformation of eq. \eqr{3.2c}.

 The entire set of equations is invariant under the transformations
 generated by
\begin{equation}
\ds{X=\xi(x)\partial_x+\eta(y)\partial_y+(\xi_x+\eta_y)\sum_{n=1}^{\infty}\bigg{
(}n-\frac{1}{2}}
\bigg{)}
\partial_{u_n}, 
\label{3.8}
\end{equation}
The surviving symmetries of the corresponding Toda lattice are 
$P$ and $D-U$ (see eq. \eqr{3.3}).

\noindent{\bf 4. The Cartan $D$ series}

In this case, the first two equations must be distinguished. They are
\alpheqn
\begin{eqnarray}
&\ds{u_{1,xy}=e^{-u_1-u_2}-e^{u_1-u_2},} \label{3.10a}\\
&\ds{u_{2,xy}=e^{-u_1-u_2}+e^{u_1-u_2}-e^{u_2-u_3}.} \label{3.10b}
\end{eqnarray}
\reseteqn
\addtocounter{equation}{-1}\refstepcounter{equation}\label{3.10}
The surviving
algebra is given by
\begin{equation}
X=\xi(x)\partial_x+\eta(y)\partial_y+(\xi_x+\eta_y)
\sum_{n=1}^{\infty}(n-1)\partial_{u_n}.
\label{3.11}
\end{equation}
Reducing further to the corresponding Toda lattice, we find that it is
invariant under translations and dilations generated by $P$ and
$D-2U$ with $P$, $D$ and $U$ as in eq. \eqr{3.3}.

Without presenting the proof \cite{1} we state that the above symmetries,
inherited from the infinite case, represent the entire symmetry
algebra in the semi-infinite case.

\subsection{The semi-infinite Toda field theories of type II related to simple 
Lie
  algebras}

The infinite system is given by eq. \eqr{1.1} with
$H_{nm}=\delta_{nm}$ and $K$ a Cartan matrix. In other words the
infinite system is
\begin{equation}
u_{n,xy}=-e^{u_{n-1}}+2e^{u_n}-e^{u_{n+1}}
\label{3.13}
\end{equation}
and the entire symmetry algebra is generated by
\begin{equation}
X=\xi(x)\partial_x+\eta(y)\partial_y
-(\xi_x+\eta_y)\sum_{n=-\infty}^{\infty}\partial_{u_n}.
\label{3.14}
\end{equation}
The gauge transformations discussed in Section 2 are all absent since
we have $p_2=p_1$. It was also shown in Section 2 that the existence
of conformal invariance depends only on the matrix $H$ which in the
field theory \eqr{1.4} is an (infinite) identity matrix. The
$A$, $B$, $C$ and $D$ Cartan series have different matrices $K$ but
this has no influence on conformal invariance. Hence the infinite,
semi-infinite and finite theories all have the same symmetry algebra
\eqr{3.14}. Moreover the same symmetries will exist even if $K$ is not
a Cartan matrix.

The reduction to the corresponding Toda lattice yields the symmetries
\begin{equation}
P=\partial_t,\quad D=t\partial_t-2\sum_{n=1}^{\infty}\partial_{u_n}
\label{3.13s}
\end{equation}
in agreement with eq. \eqr{2.8}.

\subsection{The semi-infinite field theories of type III related to
  simple Lie algebras}

We are now restricting the matrix $K$ to satisfy $K_{nm}=\delta_{nm}$
and $H$ to be a Cartan matrix. Thus, in the infinite case, the
equations we are studying are
\begin{equation}
u_{n,xy}=e^{-u_{n-1}+2u_n-u_{n+1}}.
\label{3.15}
\end{equation} 
The symmetry algebra in this case is generated by
\alpheqn
\begin{eqnarray}
X=\xi(x)\partial_x+\eta(y)\partial_y+\frac{1}{2}(\xi_x+\eta_y)\sum_{n=-\infty}^{
\infty}n^2\partial_{u_n},
\label{3.16a}
\\
V=\sum_{n=-\infty}^{\infty}\left(r_1(x)+s_1(y)+n(r_2(x)+s_2(y))\right)\partial_{
u
_n},
\label{3.16b}
\end{eqnarray}
\reseteqn
\addtocounter{equation}{-1}\refstepcounter{equation}\label{3.16}
where $\xi$, $\eta$, $r_1$, $r_2$, $s_1$, and $s_2$ are arbitrary
functions.

In the case of the corresponding infinite lattice the symmetry
algebra is generated by
\alpheqn
\begin{eqnarray}
&\ds{P=\partial_t,} &
\ds{D=t\partial_t+\sum_{n=-\infty}^{\infty}n^2\partial_{u_n},}
\label{3.17a}\\
&\ds{U_1=\sum_{n=-\infty}^{\infty}\partial_{u_n},}&
\ds{W_1=t\sum_{n=-\infty}^{\infty}\partial_{u_n}}
\label{3.17b}
\\
&\ds{U_2=\sum_{n=-\infty}^{\infty}n\partial_{u_n},}&
\ds{W_2=t\sum_{n=-\infty}^{\infty}n\partial_{u_n}.}
\label{3.17c}
\end{eqnarray}
\reseteqn
\addtocounter{equation}{-1}\refstepcounter{equation}\label{3.17}

Now let us look at individual semi-infinite cases.

\noindent{\bf 1. The Cartan $A$ series}

The first equation is
\begin{equation}
u_{1,xy}=e^{2u_1-u_2}.
\label{3.18}
\end{equation}
The others are as in \eqr{3.15}. Requiring that eq. \eqr{3.18} also be
annihilited by the symmetry algebra, we obtain the constraint
$r_1+s_1=0$. Thus the surviving symmetry algebra is the same as in
eq. \eqr{3.16} but with $r_1=s_1=0$.

For the $A$ Toda lattice, only $P$, $D$, $U_2$ and $W_2$ in \eqr{3.17}
survive.

{\bf 2. The Cartan $B$ series}

In this case the two first equations are modified. They are
\begin{equation}
\begin{array}{c}
\ds{u_{1,xy}=e^{2u_1-u_2},}\\
\ds{u_{2,xy}=e^{-2u_1+2u_2-u_3}.}
\end{array}
\label{3.19}
\end{equation}

In this case only conformal invariance survives the reduction and
takes the form
\begin{equation}
X=\xi(x)\partial_x+\eta(y)\partial_y+\frac{1}{2}(\xi_x+\eta_y)
\sum_{n=1}^\infty n(n-1)
\partial_{u_n}.
\label{3.20}
\end{equation}
However a direct calculation of the symmetry algebra shows that there
is a further gauge symmetry given by
\begin{equation}
V=(r(x)+s(y))\sum_{n=1}^{\infty}a_n\partial_{u_n},\quad
a_1=1,\;a_k=2\quad {\mbox for }\quad k\geq 2.
\label{3.21}
\end{equation}

Similarly for the Toda lattice in this case the only inherited
symmetries from \eqr{3.17} are $P$ and $D-U_2$. From a direct
calculation we obtain an additional gauge symmetry
$W=(ct+d)\sum_{n=1}^{\infty}a_n\partial_{u_n}$ with $a_n$ as in \eqr{3.21} and 
$c$ and 
$d$ constants.

\noindent{\bf 3. The Cartan $C$ series}

Only the first equation is modified and is
\begin{equation}
u_{1,xy}=e^{2u_1-2u_2}.
\label{3.22}
\end{equation}
The only surviving symmetries from eq. \eqr{3.16} are given by
\begin{equation}
X=\xi(x)\partial_x+\eta(y)\partial_y+\frac{1}{2}(\xi_x+\eta_y)
\sum_{n=1}^\infty n(n-2)
\partial_{u_n}
\label{3.23}
\end{equation}
and gauge transformations generated by vector fields of the form \eqr{3.16b} 
with 
$r_2=s_2=0$.

Similarly for the Toda lattice in this case the symmetries inherited
from \eqr{3.17} are 
\begin{equation}
\begin{array}{cc}
\ds{P=\partial_t,}&\ds{D=t\partial_t+\sum_{n=-\infty}^{\infty}n(n-2)\partial_{u
_n}
}\\ \\
\ds{U_1=\sum_{n=-\infty}^{\infty}\partial_{u_n},}&
\ds{W_1=t\sum_{n=-\infty}^{\infty}\partial_{u_n}}
\end{array}
\end{equation}

{\noindent}{\bf 2. The Cartan $D$ series}

In this case the two first equations are modified. They are
\begin{equation}
\begin{array}{c}
\ds{u_{1,xy}=e^{2u_1-u_3},}\\
\ds{u_{2,xy}=e^{2u_2-u_3}.}
\end{array}
\label{3.24}
\end{equation}

In this case only conformal invariance survives the reduction and
takes the form
\begin{equation}
X=\xi(x)\partial_x+\eta(y)\partial_y+\frac{1}{2}(\xi_x+\eta_y)
\sum_{n=1}^\infty (n^2-3n+2)
\partial_{u_n}.
\label{3.25}
\end{equation}
However a direct calculation of the symmetry algebra shows that there
is a further gauge symmetry given by
\begin{equation}
V=(r(x)+s(y))\sum_{n=1}^{\infty}a_n\partial_{u_n},\quad
a_1=a_2=1,\;a_k=2\quad {\mbox for }\quad k\geq 3.
\label{3.26}
\end{equation}

Similarly for the Toda lattice in this case the only inherited
symmetries from \eqr{3.17} are $P$ and $D-3U_2+2U_1$. From a direct
calculation we obtain an additional gauge symmetry
$W=(ct+d)\sum_{n=1}^{\infty}a_n\partial_{u_n}$ with $a_n$ as in \eqr{3.26} and 
$c$ and 
$d$ constants. 

\section{Reduction to finite Toda systems}

We will now calculate subgroups of symmetry groups of infinite Toda systems 
inherited by finite 
ones. The 
reduction procedure is explained in the Introduction. It must then be verified 
that the inherited 
groups 
indeed correspond to the complete symmetry groups of the finite systems.

\subsection{Finite Toda theories of type I}

The symmetries of the  semi-infinite Toda field theories and Toda 
lattices corresponding to $A$, $B$,
$C$ and $D$ Cartan series were 
established in Section 3.1 above. 

The corresponding finite systems are obtained by setting $u_k=0$, $k\geq N+1$ 
and imposing 
the equation
\begin{equation}
u_{N,xy}=e^{u_{N-1}-u_N},\quad{\mbox or}\quad u_{N,tt}=e^{u_{N-1}-u_N}
\label{4.1}
\end{equation}
respectively. It is easy to check that eq. \eqr{4.1} is invariant under the 
entire algebra 
\eqr{3.2}, or \eqr{3.3}, respectively (with all summations in the range $1\leq n 
\leq N$).

We obtain the following result.

\begin{theorem}
The symmetries of the finite $A_N$, $B_N$, $C_N$ and $D_N$ Toda field theories 
of type 
\eqr{1.3} are all inherited from the symmetries \eqr{3.2} of the infinite 
theories and 
coincide with those of the corresponding semi-infinite ones.
The same is true for the $A_N$, $B_N$, $C_N$ and $D_N$ Toda lattices.
\end{theorem}

\subsection{Finite Toda theories of type II}

The equations under consideration have the form \eqr{3.13}. However we modify 
the matrix 
$K$ in eq. \eqr{1.4}, the symmetries remain the same, namely conformal 
invariance 
\eqr{3.14} for Toda field theories and translations $P=\partial_t$ and 
dilations 
$D=t\partial_t-2\sum_{n=1}^N 
\partial_{u_n}$ for Toda lattices. Thus, the symmetries are the same in the 
infinite case 
and in all semi-infinite and finite cases.

\subsection{Finite Toda theories of type III}

The finite case is obtained from the semi-infinite one of Section 3.3 by setting 
$u_n=0$, 
$n\geq N+1$ and imposing
\begin{equation}
u_{N,xy}=e^{-u_{N-1}+2u_N},\quad{\mbox or}\; u_{N,tt}=e^{-u_{N-1}+2u_N},
\label{4.2}
\end{equation}
respectively. The requirement that eq. \eqr{4.2} be invariant will restrict the 
symmetry 
algebras of Section 3.3.

Let us consider individual cases.

\noindent{\bf 1. The Cartan series $A_N$}

In the $A_{1/2\infty}$ case we have the prolonged symmetry vector
\begin{equation*}
\mbox{pr}\hat{v}=
\xi(x)\partial_x+\eta(y)\partial_y+\left[\frac{1}{2}(\xi_x+\eta_y)n^2 
+(r(x)+s(y))n\right]\partial_{u_n} - (\xi_x+\eta_y)u_{n,xy}\partial_{u_{n,xy}}.
\end{equation*}
This will annihilate the eq. \eqr{4.2} only if
\begin{equation}
r(x)+s(y)=-\frac{N+1}{2}(\xi_x+\eta_y).
\label{4.3}
\end{equation}
Thus, the inherited symmetry algebra of the finite $A_N$ Toda field theory 
consists of the 
conformal transformation generated by
\begin{equation}
X=\xi(x)\partial_x+\eta(y)\partial_y+\frac{1}{2}(\xi_x+\eta_y)\sum_{n=1}^N 
n(n-N-1)\partial_{u_n}.
\label{4.4}
\end{equation}
Reducing to the $A_N$ Toda lattice we have
\begin{equation}
P=\partial_t,\quad D=t\partial_t+\sum_{n=1}^N n(n-N-1)\partial_{u_n}
\label{4.5}
\end{equation}

\noindent{\bf 2. The Cartan series $B_N$}

We require that the algebra \eqr{3.20}, \eqr{3.21} should annihilate eq. 
\eqr{4.2} (on its 
solution set). This requires $r+s=-\frac{1}{4} N(N+1)(\xi_x+\eta_y)$ and 
restricts the 
symmetry algebra to
\begin{equation}
X=\xi(x)\partial_x+\eta(y)\partial_y+\frac{1}{4}(\xi_x+\eta_y)
\left[-N(N+1)\partial_{u_1}+2\sum_{n=2}^N[n(n-1)-N(N+1)]\partial_{u_n}\right].
\label{4.6}
\end{equation}
For the Toda lattice this reduces further to
\begin{equation}
P=\partial_t,\quad D=t\partial_t+\frac{1}{4}
\left[-N(N+1)\partial_{u_1}+2\sum_{n=2}^N[n(n-1)-N(N+1)]\partial_{u_n}\right].
\label{4.7}
\end{equation}

\noindent{\bf 3. The Cartan series $C_N$}

Starting from the symmetries \eqr{3.23} and the remaining gauge 
transformations, 
we find that the symmetries of the $C_N$ Toda field theories 
reduce to conformal 
transformations, realized as
\begin{equation}
X=\xi(x)\partial_x+\eta(y)\partial_y+\frac{1}{2}(\xi_x+\eta_y)
\sum_{n=1}^N[n(n-2)-N^2+1)]\partial_{u_n}.
\label{4.8}
\end{equation}
For the $C_N$ Toda lattice this reduces to
\begin{equation}
P=\partial_t,\quad D=t\partial_t+\sum_{n=1}^N[n(n-2)-N^2+1]\partial_{u_n}.
\label{4.9}
\end{equation}

\noindent{\bf 4. The Cartan series $D_N$}

Starting from the symmetries \eqr{3.25} and \eqr{3.26} we find that eq. 
\eqr{4.2} reduces 
the symmetries to

\parbox{12cm}{
\begin{eqnarray*}
X=&\ds{\xi(x)\partial_x+\eta(y)\partial_y+\frac{1}{4}(\xi_x+\eta_y)
\bigg[-N(N-1)(\partial_{u_1}+\partial_{u_2})}
 \\
&\ds{+
2\sum_{n=3}^N[(n-2)(n-1)-N(N-1)]\partial_{u_n}\bigg]}
\end{eqnarray*}}
\parbox{1cm}{\begin{eqnarray}\label{4.10}\end{eqnarray}}
and respectively

\parbox{12cm}{
\begin{eqnarray*}
P=\partial_t,
&\ds{D=t\partial_t+\frac{1}{2}
\bigg[-N(N-1)(\partial_{u_1}+\partial_{u_2})}\\
&\ds{+
2\sum_{n=3}^N[(n-2)(n-1)-N(N-1)]\partial_{u_n}\bigg].}
\end{eqnarray*}}
\parbox{1cm}{\begin{eqnarray}\label{4.11}\end{eqnarray}}

The result is finally quite simple, namely:

\begin{theorem}
Toda field theories of the type \eqr{1.5} for a finite number $N$ of fields, 
based on the 
Cartan algebras $A_N$, $B_N$, $C_N$ and $D_N$ are invariant under conformal 
transformations only. They are realized by the vector fields \eqr{4.4}, 
\eqr{4.6}, 
\eqr{4.8} and \eqr{4.10} for $A_N$, $B_N$, $C_N$ and $D_N$, respectively. The 
symmetries 
are inherited from the semi-infinite case (however, for the $B$ and $D$ series 
the 
symmetries in the semi-infinite are not all inherited from the infinite case). 
The 
finite Toda 
lattices of type \eqr{1.5} are invariant only under a translation, and the 
appropriate 
dilations.
\end{theorem}

\section{Reduction to periodic Toda systems}

We will now study reductions from infinite (generalized) Toda systems to 
periodic ones using the 
procedure 
explained in the Introduction.  In order to be able to impose 
periodicity \eqr{per} and enlarge the symmetry algebra of the infinite system by 
adding the vector 
field 
\eqr{N},
a further condition is necessary, namely that the recursion relations \eqr{2.5} 
and 
\eqr{2.6} should have constant coefficients:
\begin{equation}
h_\sigma (n)=h_\sigma(n+1),\quad k_\sigma(m)=k_\sigma(m+1).
\label{5.2}
\end{equation}

Let us look at individual cases.

\subsection{Periodic Toda systems of type I}

 The 
inherited symmetry algebra of the usual periodic Toda field theory \eqr{1.3} is 
given by

\parbox{12cm}{
\begin{eqnarray*}
&\ds{L=x\partial_x-y\partial_y,\quad P_1=\partial_x,\quad P_2=\partial_y,}\\
&\ds{V=[\beta(x)+\gamma(y)]\sum_{n=1}^{N-1}\partial{u_n},\quad 
Z=\sum_{m=1}^{N-1}u_m\sum_{n=1}^{N-1}\partial_{u_n}.}  
\end{eqnarray*}
}
\parbox{1cm}{\begin{eqnarray}\label{5.6}\end{eqnarray}}

\noindent We see that the infinite dimensional conformal algebra is reduced to 
the Poincar\'e one.
For the Toda lattice the symmetry algebra \eqr{3.3}  
in the periodic case is reduced to $\{P,U,W,Z\}$.

The  periodic Toda field theory \eqr{1.3} can be written as eq. \eqr{1.1} with
\begin{equation}
K=
\left(
\begin{array}{rrrrr}
1&-1&0&\dots&0\\
0&1&-1&\dots&0\\
 \vdots& &\ddots&\ddots&\\
0&0&\dots&1&-1\\
-1&0&\dots&0&1
\end{array}
\right),\quad
H=
\left(
\begin{array}{rrrrr}
-1&0&0&\dots&1\\
1&-1&0&\dots&0\\
0&1&-1&\dots&0\\
 \vdots& &\ddots&\ddots&\\
0&0&\dots&1&-1
\end{array}
\right).
\label{5.7}
\end{equation}
The vector ${\bf \bar{1}}_N=(1,1,\dots,1)$ is not in the image of $H$, the 
kernel 
of $H$ and $K^T$ are one-dimensional. From Theorem 2 of Ref. \cite{1} we 
conclude that the periodic Toda field theories and lattices of type \eqr{1.3} 
have no further symmetries: all symmetries are inherited.

\subsection{Periodic Toda systems of type II}

All elements of the symmetry algebra \eqr{3.14} commute with $\hat{N}$ of eq. 
\eqr{N}. Hence, in this case the symmetry algebra is the same in the periodic 
case as in the infinite one (and also the semi-infinite and all finite ones). 
Thus, the corresponding Toda field theory is conformally invariant, the Toda 
lattice is invariant under translations $P$ and
dilations $D$ as in eq. \eqr{3.13s}. No new 
symmetries, due to the reduction, arise.

\subsection{Periodic Toda systems of type III}

The 
inherited symmetries from eq. \eqr{3.16} in this case are Poincar\'e and gauge 
invariance:

\parbox{12cm}{
\begin{eqnarray*}
&\ds{L=x\partial_x-y\partial_y,\quad P_1=\partial_x,\quad P_2=\partial_y,}\\
&\ds{V=[\beta(x)+\gamma(y)]\sum_{n=1}^{N-1}\partial{u_n}.}  
\end{eqnarray*}
}
\parbox{1cm}{\begin{eqnarray}\label{5.8}\end{eqnarray}}

\noindent For the corresponding periodic Toda lattice the only inherited 
symmetries are
\begin{equation}
P=\partial_t,\quad U=\partial_{u_n},\quad W=t\partial_{u_n}.
\label{5.9}
\end{equation}
These are the only symmetries of the system.

\section{Symmetry reductions involving continuous and discrete variables}

In this Section, we will extend the symmetry algebras of the infinite Toda 
lattices by the generator $\hat{N}$ 
given in eq. \eqr{N}. It generates transformations of the independent variable $n$ 
given by
\begin{equation}
n'=n+N.
\label{2}
\end{equation}
The quantity $N$ is to be viewed as a discrete group parameter. We can then act 
with $\hat{N}$ as if it were an element of the symmetry algebra and calculate 
group transformations and invariants in exactly the same manner as for 
differential equations.

\subsection{The infinite Toda lattice of type I}

The symmetry algebra in this case is given by eq. \eqr{3.2}. All the possible 
reductions have been studied in \cite{4}. Let us look at the interesting 
case of the reduction by the generator
\begin{equation}
\partial_n+a\partial_t,
\label{3}
\end{equation}
where $a$ is a constant \cite{4}. The invariants are $\xi=an-t$ and $u_n$. We thus 
consider solutions of the infinite Toda lattice of the form 
\begin{equation}
u_n(t)=F(\xi),\;\;\xi=an-t.
\label{31/2}
\end{equation}
Substituing into \eqr{1.3} we find the following equation for $F$
\begin{equation}
F_{\xi \xi}={\mbox e}^{F(\xi-a)-F(\xi)}-{\mbox e}^{F(\xi)-F(\xi+a)}.
\label{4}
\end{equation}
A soliton solution of \eqr{4} is given by 
\begin{equation}
F(\xi)=\mbox{ln}\left(\frac{1+\exp{[2\xi \sinh{\frac{\alpha}{2}}\pm 
\alpha]}}{1+\exp{[2\xi \sinh{\frac{\alpha}{2}}]}}\right),\;\;
a=\pm \frac{\alpha}{2\sinh{\frac{\alpha}{2}}}.
\label{5}
\end{equation}

\subsection{The infinite Toda lattice of type II}

The symmetry algebra in this case is given by eq. \eqr{3.13s}. We first consider 
reductions by the generator \eqr{3}. Invariant solutions then have the form 
given by eq. \eqr{31/2} and the function $F$ satisfies
\begin{equation}
\label{6}
F_{\xi \xi}=-\mbox{e}^{F(\xi-a)}+2\mbox{e}^{F(\xi)}-\mbox{e}^{F(\xi+a)}.
\end{equation} 
We can also consider reductions by the operator $\hat{N}+aD$ with $D$ given 
in eq. \eqr{3.13s}. Invariant solutions are then of the form
\begin{equation}
u_n(t)=F(\eta)-2an,\;\;\eta=t\mbox{e}^{-an}.
\end{equation}
The function $F$ satisfies the equation
\begin{equation}
\label{7}
F_{\eta \eta}=-\mbox{e}^{F(\mbox{e}^a\eta)}+2\mbox{e}^{F(\eta)}-
\mbox{e}^{F(\mbox{e}^{-a}\eta)}.
\end{equation}

\subsection{The infinite Toda lattice of type III}

The symmetry algebra in this case is given by eq. \eqr{3.17}. We first consider 
reductions by generators of the form 
\begin{equation}
\hat{N}+aP+cU_1+dU_2+eW_1+fW_2,
\end{equation}
where $P$, $U_1$, $U_2$, $W_1$ and $W_2$ are given in \eqr{3.17}. Invariant 
solutions will have the form
\begin{equation}
\label{8}
u_n(t)=F(\xi)+(c-\xi e)n+\frac{n^2}{2}(D+ae-\xi f)+af\frac{n^3}{3},
\end{equation}
where $\xi=an-t$. The function $F$ then satisfies the equation
\begin{equation}
\label{9}
F_{\xi \xi}=\exp{[-F(\xi+a)+2F(\xi)-F(\xi-a)+ae-d+f\xi]}.
\end{equation}
In the case when $d=ae$ and $f=0$, we have a solution quadratic in $\xi$
\begin{equation}
\label{10}
F=\alpha \xi^2+\beta\xi+\gamma,
\end{equation}
where $\alpha$ is determined in terms of $a$ by the equation
\begin{equation}
2\alpha=\mbox{e}^{-2\alpha a^2}.
\end{equation}
The constants $\beta$ and $\gamma$ are free.

We also consider reductions by generators of the form
\begin{equation}
\hat{N}+bD+cU_1+dU_2+eW_1+fW_2
\end{equation}
with $b$ nonzero. In this case invariant solutions have the form
\begin{equation}
u_n(t)=b\frac{n^3}{3}+d\frac{n^2}{2}+cn+\frac{\mbox{e}^{bn}}{b^2}\eta(be+bf-f)+
F(\eta),
\end{equation}
where $\eta=t\mbox{e}^{-bn}$. The function $F$ must then satisfy the equation
\begin{equation}
F_{\eta \eta}=\exp{(d-F(\mbox{e}^b\eta)+2F(\eta)-F(\mbox{e}^{-b}\eta))}.
\end{equation}

\section{Conclusions}

The starting point of this article are the symmetries \eqr{2.2} of a large class 
of 
classical field theories with exponential interactions described by eq. 
\eqr{1.1}. The 
symmetries \eqr{2.2}, established in our earlier article \cite{1} are present 
when the 
discrete variable $n$, labeling the fields, varies in an infinite range 
$-\infty<n<\infty$. The ``interaction matrices'' $K$ and $H$ in eq. \eqr{1.1} 
are very 
general band  matrices with bands of constant width (see eq. \eqr{2.3} and 
\eqr{2.4}). 
From eq. \eqr{2.2a} we see that the theory is always conformally invariant. In 
view of 
eq. \eqr{2.2b}, the theory is gauge invariant, unless the matrix $H$ is 
diagonal. If $K$ 
is diagonal, the gauge group is abelian. If both $H$ and $K$ are non diagonal, 
the gauge 
group is nonabelian.

If the generalized Toda field theories \eqr{1.1} correspond to difference 
equations with 
constant coefficients, then the symmetry group includes a further 
transformation, namely 
translations of $n$
\begin{equation}
\tilde{n}=n+N,\quad N\in \mathbb{Z},
\end{equation} 
formally generated by the operator \eqr{N}. This occurs if the band matrices $H$ 
and 
$K$, specified in eq. \eqr{2.3} and \eqr{2.4}, satisfy condition \eqr{5.2}, that 
is, if the 
corresponding 
difference 
equations have constant ($n$-independent) coefficients.

The symmetry group of the infinite ($-\infty<n<\infty$) generalized Toda field 
theory is 
applied to study different types of reductions of the system. 

It should be emphasized that when we perform symmetry reduction, for 
differential 
equations, difference equations, or differential-difference ones, we obtain a 
symmetry 
algebra inherited from the original symmetry algebra. This may not be the entire 
symmetry 
algebra of the reduced system. A prime example for this is provided by the 
Laplace 
equation in three-dimensional euclidian space $E_3$. It is invariant under the 
conformal 
group O(4,1). If we reduce it by translational invariance to the Laplace 
equation in 
$E_2$, we obtain the inherited symmetry group O(3,1). However, the reduced 
equation is 
invariant under a much larger group, the infinite-dimensional conformal group 
$E_2$. The 
presence of this noninherited symmetry group is an indication of the fact that 
two-dimensional theories differ qualitatively from three-dimensional ones. We 
have seen 
above that no such radical difference between ``inherited symmetries'' and ``all 
symmetries'' occurs in reductions of Toda systems.

Among open problems, not addressed in this article, we mention two related ones. 
They are: 
under what conditions on the matrices $H$ and $K$ are the generalized Toda 
systems 
\eqr{1.1} and \eqr{1.7} integrable? When do these equations allow higher 
symmetries that 
depend on the derivatives of $u_n$, or that are nonlocal?

\section*{Acknowledgments}

The research reported in this article is part of a project supported by NATO 
grant CRG 960717. P.W. 
and S.L. 
acknowledges support from NSERC of Canada, and FCAR du Qu\'ebec. They also thank 
the Dipartimento 
di Fisica, 
Universit\`a di Lecce, for its hospitality.

\end{document}